\documentclass[final,onecolumn,rmp,floats,amsfonts,amssymb,amsmath,superscriptaddress,floatfix,a4paper]{revtex4}
\usepackage{graphicx}
\usepackage{hyperref}
\usepackage{subfigure}
\usepackage{url}
\usepackage{layout}

\begin{document}
%\layout
\title{Experimental studies of liquid-liquid dispersion in a turbulent shear flow.}
\author{Florent Ravelet}
\affiliation{Laboratory for Aero and Hydrodynamics, Leeghwaterstraat 21, 2628 CA Delft, The Netherlands.}
\email{florent.ravelet@ensta.org}
\author{Ren\'e Delfos Jerry Westerweel}
\affiliation{Laboratory for Aero and Hydrodynamics, Leeghwaterstraat 21, 2628 CA Delft, The Netherlands.}
\author{Jerry Westerweel}
\affiliation{Laboratory for Aero and Hydrodynamics, Leeghwaterstraat 21, 2628 CA Delft, The Netherlands.}
\maketitle

\section{Introduction}
Liquid-liquid dispersions are encountered for instance in extraction or chemical engineering when contact between two liquid phases is needed. Without surfactants, when two immiscible fluids are mechanically agitated, a dispersed state resulting from a dynamical equilibrium between break-up and coalescence of drops can be reached. Their modelling in turbulent flows is still limited \cite{portela2006}.

\subsubsection{Experimental setup}

We built up an experiment to study liquid-liquid dispersions in a turbulent Taylor-Couette flow, produced between two counterrotating coaxial cylinders, of radii $r_i=100\,$mm and $r_o=110\,$mm (gap ratio $\eta=r_i/r_o=0.909$). The useful length is $L=185\,$mm. There is a space of $10\,$mm between the cylinders bottom ends and a free surface on the top. The cylinders rotation rates $\omega_i$ and $\omega_o$ can be set independently. We use the set of parameters defined in \cite{dubrulle2005}: a mean Reynolds number $Re$ based on the shear and on the gap; and a \lq\lq Rotation number\rq\rq~$Ro$ which is zero in case of perfect counterrotation ($r_i \omega_i= - r_o \omega_o$). At maximal rotation frequency, $Re \simeq 5.6 \times 10^4$ for pure water.

For the two-phases studies, we use a low-viscous oil (Shell Macron 110), with $\mu =3 \times 10^{-3}\,$Pa.s, $\rho=800\,$kg.m$^{-3}$. The acqueous phase is either pure water or an NaI solution with a refractive index matched to that of oil ($1.445$ at $20^o$C). The viscosity, density and interfacial tension for the NaI are $\mu = 2 \times 10^{-3}\,$Pa.s, $\rho=1550\,$kg.m$^{-3}$ and $\sigma \simeq 15\,$mN.m$^{-1}$. The viscosity ratio is close to unity and the density ratio is close to $2$.  

We give here some preliminary results on thresholds for dispersion, and on the increase of wall shear stress, based on global measurements of the torque $T$, using an HBM T20WN rotating torquemeter. 

\section{Results and discussion}
\subsubsection{Turbulent Taylor-Couette flow in exact counterrotation}
\begin{figure}
\centering
\includegraphics[clip,height=4cm]{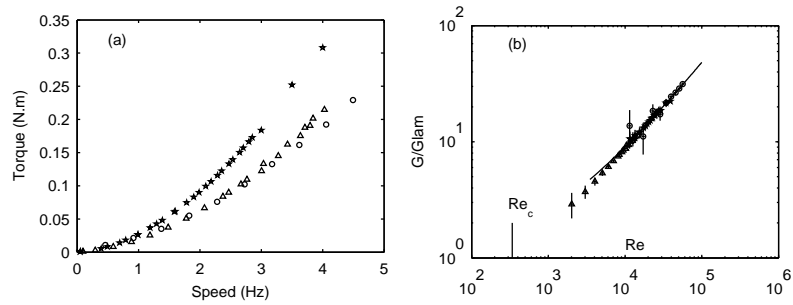}
\caption{(a): Torque {\em vs.} rotation frequency in exact counterrotation for Oil ($\triangle$), Water ($\circ$) and NaI ($\star$). (b): Dimensionless torque $G$ normalised by laminar torque $G_{lam}=2\pi \eta / (1-\eta)^2 Re$ {\em vs.} $Re$. Solid line is a fit of the form $G=a\,Re^2/(ln(b\,Re^2))^{(3/2)}$ \cite{dubrulle2002}, with $a=16.5$ and $b=7\times 10^{-5}$.}
\label{fig:1}
\end{figure}

Though the Taylor-Couette flow has been widely studied, few experimental results and theories are available for $Ro=0$ \cite{dubrulle2005}. We first report torque measurements on pure fluids in Fig.~\ref{fig:1}. The torque due to the bottom part has been removed by studying different filling levels. Esser and Grossmann \cite{esser1996} proposed a formula for the first instability threshold. Here it gives $Re_c(\eta,Ro)=338$. Moreover, Dubrulle and Hersant \cite{dubrulle2002} proposed a formula for the dimensionless torque $G=T/(\rho \nu^2 L)$ in the turbulent regime which is also consistent with our data (Fig.~\ref{fig:1}b).

\subsubsection{Increase of dissipation in presence of turbulent dispersions}

\begin{figure}
\centering
\includegraphics[clip,height=4cm]{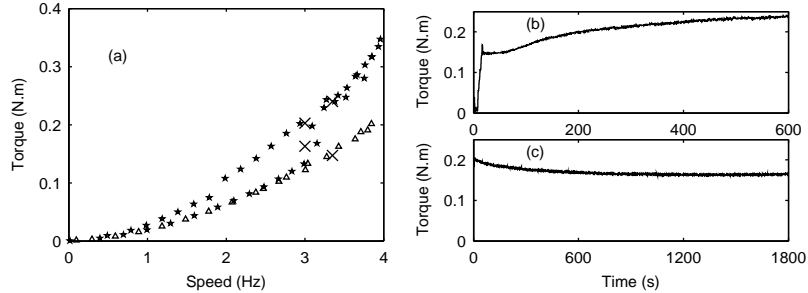}
\caption{(a): Torque vs. rotation frequency for pure Oil ($\triangle$) and $33\%$ Water in Oil ($\star$). Speed is increased by steps of $0.2\,$Hz every $60\,$s, up to $4\,$Hz and then decreased back to $0$. ($\times$) are values taken from experiments (b) and (c). (b): From a two-layer system at rest, the speed is set to $3.5\,$Hz. (c): The speed is then decreased to $3\,$Hz.}
\label{fig:2}
\end{figure}

At rest, two immiscible fluids are separated into two layers. Increasing the speed, the fluids gradually get dispersed one into another. The torque shows up a fast increase around $f=3\,$Hz (Fig.~\ref{fig:2}a), corresponding to an homogeneous dispersion. When thereafter decreasing the speed, there seems to be hysteresis in the system. In fact the time scales involved in the dispersion and separation processes are very long, the latter being even longer (Fig.~\ref{fig:2}b-c). For this volume fraction of oil and water, the two-layer system is the stable state below $2.5\,$Hz and the fully dispersed is stable above $3\,$Hz. Complicated behaviours including periodic oscillations (with a period of $20\,$min) have been observed in between and requires further investigation.

\begin{figure}
\centering
\includegraphics[height=3.5cm]{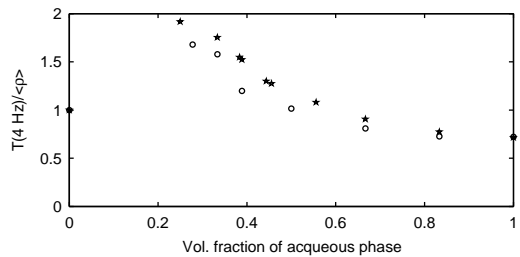}
\caption{Massic torque at $4\,$Hz {\em vs.} volume fraction of Water ($\circ$) and NaI ($\star$).}
\label{fig:3}
\end{figure}

The torque per unit mass in a dispersed state at $4\,$ Hz is maximum for a volume fraction of the acqueous phase around $33\%$ and is 2 times that of pure Oil (Fig.~\ref{fig:3}). Similar behaviours have been evidenced recently in liquid-liquid turbulent pipe flow experiments \cite{piela2006}.

\subsubsection{Perspectives}
Further investigations on the scaling for the counterrotating monophasic flow will be performed, on a wider Reynolds number range. We have already optimised the calibration procedure for Stereoscopic PIV measurements in a new setup of better optical quality. Further work is needed to understand what are the relevant parameters for dispersion to arise (Froude number, Weber number) We will then address the question of drop size distribution using Light Induced Fluorescence. First results have shown the presence of multiple drops.

\end{document}